A nonlinear quantum –dynamical system of spin ½ particles

based on the classical Sine-Gordon Equation

Yair Zarmi

Jacob Blaustein institutes for Desert Research

Ben-Gurion University of the Negev

Midreshet Ben-Gurion, 84990 Israel

Abstract

The Hirota transformation for the soliton solutions of the classical Sine-Gordon equation is suggestive of an extremely simple way for the construction of a nonlinear quantum-dynamical system of spin ½ particles that is equivalent to the classical system over the soliton sector. The soliton solution of the classical equation is mapped onto an operator, U, a nonlinear functional of the particle-number operators, that solves the classical equation. Multi-particle states in the Fock space are the eigenstates of U; the eigenvalues are the soliton solutions of the Sine-Gordon equation. The fact that solitons can have positive as well negative velocities is reflected by the characterization of particles in the Fock space by two quantum numbers: a wave number $k$, and a spin projection, $\sigma$ (= ±1). Thanks to the simplicity of the construction, incorporation of particle interactions, which induce soliton effects that do not have a classical analog, is simple.

The Sine-Gordon (SG) equation has attracted extensive attention over the years, both as a classical PDE (describing quantum mechanical phenomena) [1-6] at one extreme, and an equation for study within the framework of quantum-field theory [7-15] at the other, the quantization of the classical equation being a major issue. Invariably, quantization ends up with replacement of the solution of the classical equation by a field operator.

The question that this paper addresses is whether it is possible to construct a nonlinear quantum-dynamical system that is associated with the SG equation, and does not exploit the notion of the field operator. Examples of a possible answer to this question in the cases of the KdV [16], mKdV [17], Sawada-Kotera [18] and bidirectional KdV equations [19, 20] have been proposed in [21], where the Hirota algorithm for the construction of the soliton solutions of these equations [19, 22, 23] was used. The soliton solution of the classical equation is mapped onto an operator, U, a nonlinear functional of the particle-number operators, that solves the classical equation. Multi-particle states in the Fock space are the eigenstates of U; the eigenvalues are the soliton solutions of the evolution equation. While the proposed procedure does not constitute an alternative to the powerful traditional quantization, it opens the door to new ways of constructing nonlinear quantum-dynamical systems that are associated with classical integrable nonlinear evolution equations. Thanks to the simplicity of the construction, incorporation of particle interactions, which induce soliton effects that do not have a classical analog, is simple.

The soliton solutions of the Sine-Gordon equation,

$$\frac{\partial^2 u}{\partial x^2} - \frac{\partial^2 u}{\partial t^2} = \sin u \ , \tag{1}$$

are obtained by the following Hirota transformation [24]:

$$u(t,x;\vec{q},\vec{\sigma}) = 4\tan^{-1}\left[g(t,x;\vec{q},\vec{\sigma})\big/f(t,x;\vec{q},\vec{\sigma})\right] \ . \tag{2}$$

In Eq. (2),

$$g(t,x;\vec{q},\vec{\sigma}) = \sum_{\substack{1 \leq n \leq N \\ n\ \text{odd}}} \left( \sum_{1 \leq i_1 < \cdots < i_n \leq N} \left\{ \prod_{j=1}^{n} \varphi(q_{i_j}, \sigma_{i_j}; t, x) \prod_{i_l < i_m} V(q_{i_l}, \sigma_{i_l}, q_{i_m}, \sigma_{i_m}) \right\} \right) , \qquad (3)$$

$$f(t,x;\vec{q},\vec{\sigma}) = 1 + \sum_{\substack{2 \leq n \leq N \\ n\ \text{even}}} \left( \sum_{1 \leq i_1 < \cdots < i_n \leq N} \left\{ \prod_{j=1}^{n} \varphi(q_{i_j}, \sigma_{i_j}; t, x) \prod_{i_l < i_m} V(q_{i_l}, \sigma_{i_l}, q_{i_m}, \sigma_{i_m}) \right\} \right) , \qquad (4)$$

$$\vec{q} \equiv \{q_1, q_2, \ldots, q_N\} \qquad (q_1 < q_2 < \ldots < q_N) \qquad \vec{\sigma} \equiv \{\sigma_1, \sigma_2, \ldots, \sigma_N\} , \qquad (5)$$

$$\varphi(q,\sigma;t,x) = e^{qx - v(q,\sigma)t + \delta(q,\sigma)} , \qquad (6)$$

$$v(q,\sigma) = \sigma\sqrt{q^2 - 1} , \qquad \sigma = \pm 1 , \qquad (7)$$

and

$$V(q,\sigma,q',\sigma') = \frac{(q-q')^2 - (v(q,\sigma) - v(q',\sigma'))^2}{(q+q')^2 - (v(q,\sigma) + v(q',\sigma'))^2} = -\left( \frac{(q + v(q,\sigma)) - (q' + v(q',\sigma'))}{(q + v(q,\sigma)) + (q' + v(q',\sigma'))} \right)^2 . \qquad (8)$$

Soliton wave numbers are denoted by $q$, and the signs of soliton velocities – by $\sigma$.

The corresponding nonlinear quantum-dynamical system is constructed over a Fock space of spin-1/2 Fermions, with the functions $f(t,x;\vec{q},\vec{\sigma})$ and $g(t,x;\vec{q},\vec{\sigma})$ replaced by the operators

$$\mathrm{F}(t,x) = 1 + \sum_{\substack{\sigma_i = \pm 1 \\ \sigma_m = \pm 1}} \sum_{\substack{n=2 \\ n\ \text{even}}}^{\infty} \frac{1}{n!} \int_0^\infty \int_0^\infty \cdots \int_0^\infty \left\{ \left( \prod_{i=1}^{n} \varphi(k_i \sigma_i; t, x) \mathrm{N}_{k_i, \sigma_i} \right) \left( \prod_{1 \leq l < m \leq n} V(k_l, \sigma_l, k_m \sigma_m) \right) \right\} dk_1\, dk_2 \cdots dk_n , \qquad (9)$$

and

$$\mathrm{G}(t,x) = \sum_{\substack{\sigma_i = \pm 1 \\ \sigma_m = \pm 1}} \sum_{\substack{n=1 \\ n\ \text{odd}}}^{\infty} \frac{1}{n!} \int_0^\infty \int_0^\infty \cdots \int_0^\infty \left\{ \left( \prod_{i=1}^{n} \varphi(k_i \sigma_i; t, x) \mathrm{N}_{k_i, \sigma_i} \right) \left( \prod_{1 \leq l < m \leq n} V(k_l, \sigma_l, k_m \sigma_m) \right) \right\} dk_1\, dk_2 \cdots dk_n . \qquad (10)$$

(A regularization procedure ought to be employed in calculating matrix elements of these operators. This has been discussed in detail in [21].)

In Eqs. (9) and (10), $N_{k,\sigma}$ is the particle-number operator, expressed in terms of the particle-creation and annihilation operators over a Fock space of spin-1/2 particles:

$$N_{k,\sigma} = a^\dagger_{k,\sigma} a_{k,\sigma} \qquad \left(\{a^\dagger_{k,\sigma}, a_{k',\sigma'}\} = \delta(k-k')\delta_{\sigma,\sigma'}\right) . \qquad (11)$$

One readily finds that multi-particle states are the eigenstates of these operators:

$$F(t,x)\Psi_N = f(t,x;\vec{q},\vec{\sigma})|\Psi_N\rangle, \qquad G(t,x)\Psi_N = g(t,x;\vec{q},\vec{\sigma})|\Psi_N\rangle$$

$$\left(|\Psi_N\rangle = \left(\prod_{i=1}^{N} a^\dagger_{q_i,\sigma_i}\right)|0\rangle\right) \qquad (12)$$

where the functions $g(t,x;\vec{k},\vec{\sigma})$ and $f(t,x;\vec{k},\vec{\sigma})$ given in Eqs. (3) and (4), generate the soliton solution of Eq. (1).

The operator, $U(t,x)$, generates the soliton solutions of Eq. (1). It is defined by an operator analog of Eq. (2):

$$G(t,x)F(t,x)^{-1} = \tan\left[\frac{1}{4}U(t,x)\right] . \qquad (13)$$

(For reasons of convergence issues, the inverse of the Hirota transformation is presented in Eq. (13)).

The action of the right hand side of Eq. (13) on a state in the Fock space yields:

$$\tan\left[\frac{1}{4}U(t,x)\right]|\Psi_N\rangle = \frac{g(t,x;\vec{k},\vec{\sigma})}{f(t,x;\vec{k},\vec{\sigma})}|\Psi_N\rangle . \qquad (14)$$

Therefore, $u(t,x)$ of Eq. (2) is the eigenvalue of the operator $U(t,x)$. Moreover, as $U(t,x)$ is a diagonal operator, it obeys the classical Sine-Gordon equation:

$$\frac{\partial^2 U}{\partial x^2} - \frac{\partial^2 U}{\partial t^2} = \sin U \quad . \tag{15}$$

As a result, the classical Hamiltonian that generates the Sine-Gordon equation becomes the Hamiltonian for the quantum system if $u$ is replaced by U:

$$H = \int_{-\infty}^{+\infty}\left\{\tfrac{1}{2}(u_t)^2 - \tfrac{1}{2}(u_x)^2 + (1-\cos u)\right\}dx \Rightarrow \int_{-\infty}^{+\infty}\left\{\tfrac{1}{2}(U_t)^2 - \tfrac{1}{2}(U_x)^2 + (1-\cos U)\right\}dx \quad . \tag{16}$$

If a perturbation is added to the classical equation, Eq. (1), then, most often, the solution of the perturbed equation must be obtained through a perturbation expansion. Usually, the perturbation series is a divergent asymptotic series that provides an order-by-order correction to the zero-order approximation. The higher order corrections are functionals of the zero-order approximation, and, as such, do not have the capacity to change the identity of the solitons generated by the zero-order approximation. The situation is quite different in the new quantum-dynamical system. The simple construction easily allows for the incorporation of perturbations that affect the solitons in a manner that does not have a classical analog. Consider the following perturbed Eq. (1):

$$\frac{\partial^2 u}{\partial x^2} - \frac{\partial^2 u}{\partial t^2} = \sin u + \varepsilon\left(u\cos u - \sin u\right) \quad . \tag{17}$$

The solution through $O(\varepsilon)$ is:

$$u = (1-\varepsilon)u_0 \quad . \tag{18}$$

A possible quantized version of Eq. (17) is:

$$\frac{\partial^2 U}{\partial x^2} - \frac{\partial^2 U}{\partial t^2} = \sin U + \varepsilon\, P\left(U\cos U - \sin U\right) \quad . \tag{19}$$

In Eq. (19), choose the operator, P, to have the capacity to create an annihilate particles:

$$P= \sum_{\sigma,\sigma'=\pm 1} \int_{-\infty}^{+\infty} \int_{-\infty}^{+\infty} c(k,\sigma,k',\sigma') a_{k,\sigma}^{\dagger} a_{k',\sigma'} \, dk \, dk'$$
$$\left( \int_{-\infty}^{+\infty} \int_{-\infty}^{+\infty} |c(k,\sigma,k',\sigma')|^2 \, dk \, dk' < \infty \right) \quad . \tag{20}$$

Through $O(\varepsilon)$, Eq. (19) is solved by

$$U = (1 - \varepsilon P) U_0 \quad . \tag{19}$$

The zero-order term, $U_0$, obeys Eq. (15), hence, can be constructed through Eqs. (12) and (13). As a result, diagonal matrix elements of the perturbed solution, Eq. (19), will be similar to the classical expression:

$$\langle \Psi_N | U | \Psi_N \rangle = \left( 1 - \varepsilon \sum_{i=1}^{N} c(q_i,\sigma_i,q_i,\sigma_i) \right) u_N(t,x;\vec{q},\vec{\sigma}) \quad . \tag{20}$$

In Eq. (20), $u_N(t,x;\vec{q},\vec{\sigma})$ denotes an $N$-soliton solution, with wave-number and spin-projection vectors $\vec{k}$ and $\vec{\sigma}$, respectively. However, off-diagonal matrix elements will mix in other solitons. For example, in the single-particle subspace:

$$\langle q,\sigma | U | q',\sigma' \rangle = \varepsilon c(q,\sigma,q',\sigma') u(t,x;q',\sigma') \quad . \tag{21}$$

Thus, the eigenvalues of U will contain higher-order corrections that mix in other solitons.

Acknowledgment The author wishes to thank S.N.M. Ruijsenaars for pointing out the interest in the application to the Sine-Gordon equation.